\documentclass[3p,times]{elsarticle}

\usepackage{ecrc}

\volume{00}

\firstpage{1}

\journalname{Nuclear Physics A}

\runauth{}

\jid{nupha}

\usepackage{amssymb}

\usepackage[figuresright]{rotating}

\begin{document}

\begin{frontmatter}



\dochead{}

\title{Two particle correlation measurements at PHENIX}


\author{T. Todoroki\corref{email} for the PHENIX Collaboration}
\cortext[email]{E-mail address: todoroki@rcf.rhic.bnl.gov}

\address{University of Tsukuba,Tsukuba, Ibaraki 305, Japan\\
 RIKEN Nishina Center for Accelerator-Based Science, Wako, Saitama 351-0198, Japan}

\begin{abstract}
Measurements of two particle azimuthal correlations in relativistic heavy ion collisions
are useful tools to dissect the interplay between hard-scattered partons and hot dense medium.
Correlations with trigger particle selection relative to second order event plane are sensitive to
the path-length dependence of parton energy loss and to the influence of the medium on jet
for high and intermediate transverse momenta pairs, respectively.
To study the parton-medium coupling, it is also crucial to obtain correlations with rejection 
of contributions from higher harmonic flow. We present current results of second order event plane
dependent correlations as well as correlations in which contributions from higher harmonic flow
have been excluded in Au+Au collisions at $\sqrt{s_{NN}}=200$ GeV measured by PHENIX.
\end{abstract}


\end{frontmatter}


\section{Higher harmonic event planes and flow}
\label{higher}
In previous understanding of relativistic heavy ion collisions, it has been assumed that the spatial
geometry of the participant nucleons shows almond-like shapes and considered that the generated 
hot and dense medium mainly expands in the direction of short axis of almond-like shapes.

Recent studies with AMPT model simulations\cite{Alver:2010prc81} and experimental measurements
\cite{Adare:2011prl107}\cite{Aamodt:2011prl107}\cite{Aad:2012prc86}
revealed that, in addition to ellipticity, there are triangularity, squarity as well as
higher-order deformation in the initial collision geometry originating from its fluctuations,
giving thus rise to higher-order flow harmonics, i.e. anisotropy of particle productions relative to
each harmonic event plane, over wide rapidity range. The effects due to fluctuations of initial collision
geometry play an important roll for the understanding of space time evolution of the medium created
in heavy ion collisions.

The amplitude of flow harmonics $v_n$ is defined as  
\begin{eqnarray}
& & \frac{dN}{d\phi}
\propto
1 + \sum_{n=1} 2 v_{n} \cos (n(\phi-\Phi_{n})),
\nonumber \\
& &
v_n = \left< \cos ( n (\phi - \Phi_{n}) ) \right> , (n=1,2,3,...),
\label{eq:vn}
\end{eqnarray}
where $\phi$ is the particle azimuthal angle and $\Phi_n$ is the direction of the nth harmonic event plane 
in transverse plane.

The $v_n$ of hadrons within $|\eta|<0.35$ is measured with the event planes determined in $1.0<\eta<2.8$ to ensure
a sufficient pseudo-rapidity gap between particle and event planes to reduce non-flow correlations,
such as those due to jet production.

\section{Two particle correlations}
\label{two-particle}
Two particle correlations are calculated by dividing the relative angular distributions of real event pairs
with those from mixed events, and applying proper normalization:
\begin{eqnarray}
& &C(\Delta\phi)=\frac{N^{real}_{pair}(\Delta\phi)}{N^{mix}_{pair}(\Delta\phi)}\frac{\int\Delta\phi N^{mix}_{pair}(\Delta\phi)}{\int\Delta\phi N^{real}_{pair}(\Delta\phi)},\\
& &
\Delta\phi=\phi^{asso.}-\phi^{trig.},
\label{eq:two-particle}
\end{eqnarray}
where $\phi^{trig}$ is the trigger particle azimuthal angle and $\phi^{asso}$ is the associate particle one.
Correlations have been measured with particles tracked at mid rapidity in the range $|\eta|<0.35$
without rapidity gaps between trigger and associate particles. In the correlations, the contributions from jet survives,
thus allowing us to study to the interplay between hard-scattered partons and medium.

In addition to inclusive trigger correlations, the study of two particle correlations with additional selection 
of the azimuthal angle between the trigger particle and the second order event plane provides a useful tool
to control the path length which parton propagate. This studies produce measurements sensitive to the path-length 
dependence of parton energy loss with high momenta pairs and to the influence of medium effects with intermediate 
momenta pairs.

The correlations include the contributions from collective flow. The flow shapes can be estimated
from experimentally measured event plane resolutions and the Fourier coefficients $v_n$.
Their contribution is subtracted with the normalization factor determined by ZYAM method\cite{Adler:2006prl97}.

\section{Results}
\label{results}

\subsection{Two-particle correlations versus event plane at high pt}
\begin{figure}[tbp]
\begin{center}
\includegraphics[scale=0.65]{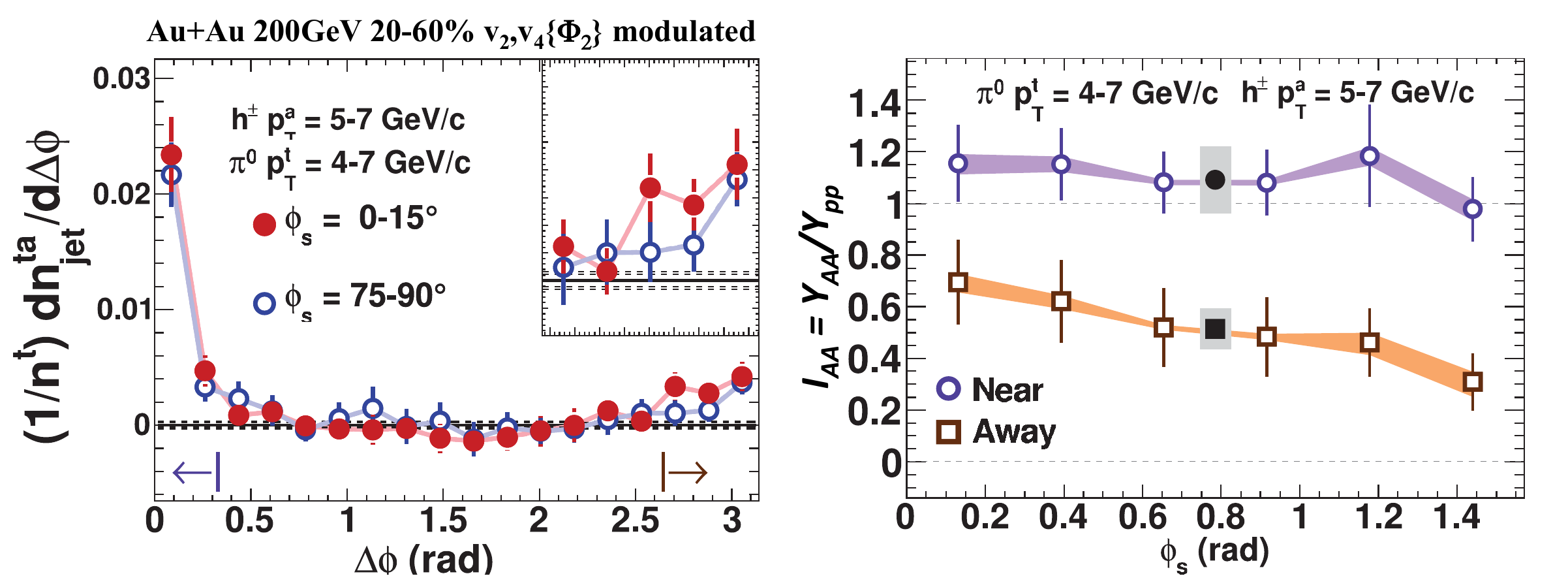}
\end{center}
\caption{
(Color online)
$\pi^0$-hadron azimuthal correlations in Au+Au collisions at $\sqrt{s_{NN}}=$200 GeV in centrality 20-60\%\cite{Adare:2011prc84}.
Transverse momentum ranges are 4-7 GeV/c for trigger $\pi^0$ and 5-7 GeV/c for associated hadrons.
Contributions from $v_2$ and $v_4(\Phi_2)$ are subtracted with ZYAM method. Left panel shows triggered 
correlations for the most in-plane and most out-of-plane intervals.
Right panel shows the near and away side correlation yield as a functions of trigger particle angle
relative to second order event plane $\phi_s = \phi^{trig} - \Phi_2$ normalized by correlation yield
in $p+p$ collisions.
}
\label{fig:2pchigh}
\end{figure}

The $\pi^0$-hadron correlations with neutral pions in the $p_T$ range $4<p_T^{\pi^0}<7$ GeV/c
as trigger particles and hadrons in the range $5<p_T^{h^\pm}<7$ GeV/c as associated particles
are shown in the left panel of Fig. 1 for Au+Au collisions at $\sqrt{s_{NN}}=200$ GeV and
in the centrality range 20-60\%, where path length can be well controlled.
Two sets of data points are shown, corresponding to different intervals of the angle between
the trigger particle and the second order event plane.

Right panel of Fig.\ref{fig:2pchigh} shows the integrated yield in near and away side of triggered
correlations, normalized to the yields in $p+p$ collisions as a functions of
the relative angle between the trigger particle ($\pi^0$) and the second order event plane ($\Phi_2$).
Here the collective flow contributions from $v_2$ and $v_4(\Phi_2)$ harmonics are subtracted,
while $v_3$ and $v_4$ are not subtracted.
However the omission of higher $v_n$ is not significant in correlation measurements with high momenta pairs,
since the ratio of jet signal over flow contribution is large enough.

While the near side yield is consistent with the yield in $p+p$ collisions within the statistical
and systematic uncertainties, the away side yield is significantly suppressed compared to $p+p$ collisions.
The away side yield is monotonically decreasing as the angle between the trigger particle
and the second order event plane increase, that is parton path length increases. 
\\

\subsection{Two-particle correlations versus event plane at intermediate $p_T$}
\begin{figure}[tbp]
\begin{center}
\includegraphics[scale=0.60]{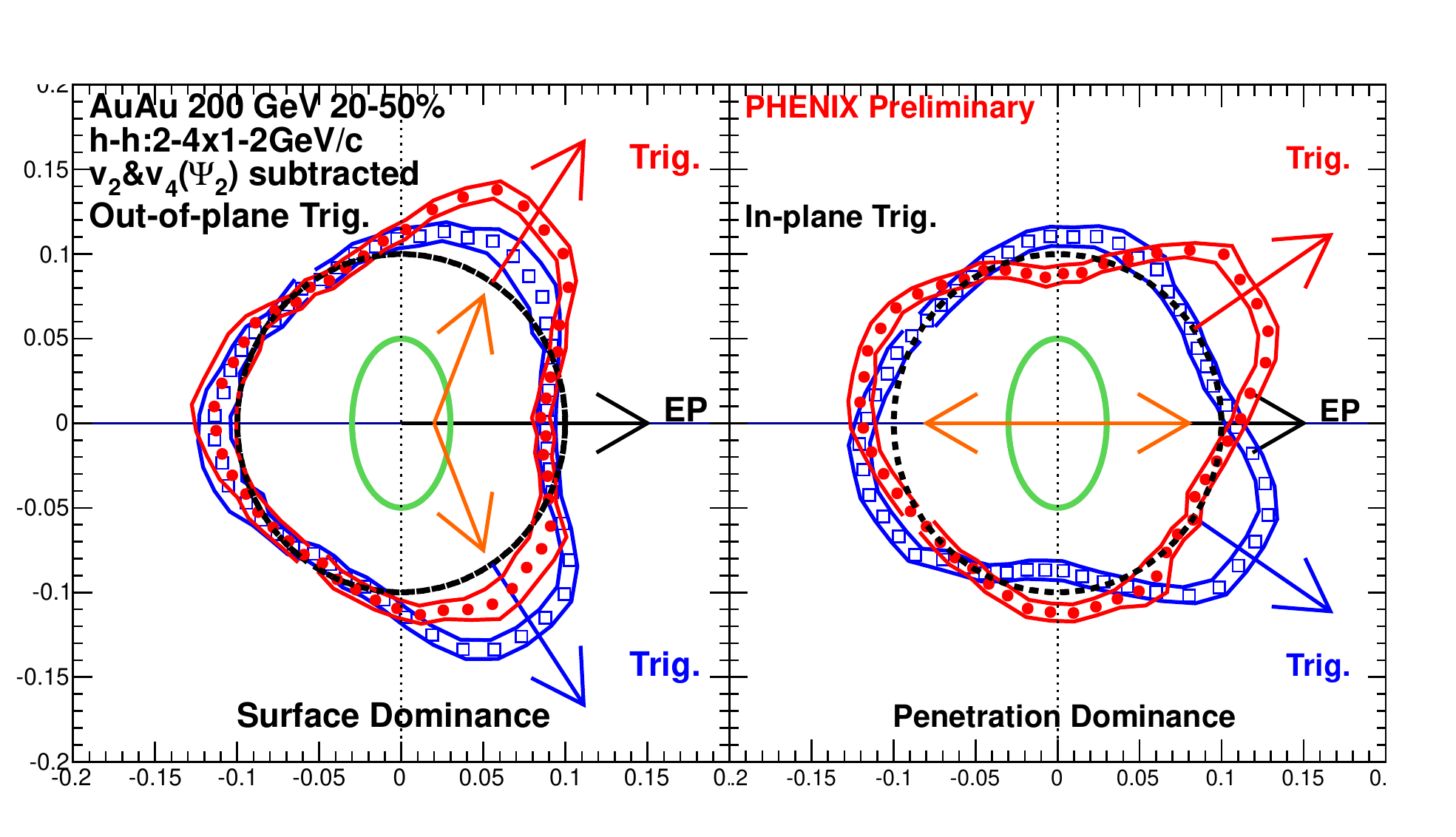}
\end{center}
\caption{
(Color online)
Two particle charged hadron correlations in Au+Au collisions at $\sqrt{s_{NN}}=$200 GeV in centrality 20-50\%.
Transverse momentum ranges are 2-4 GeV/c for trigger particles and 1-2GeV/c for associated ones.
Contributions from $v_2$ and $v_4(\Phi_2)$ are subtracted with ZYAM method.
}
\label{fig:2pcint}
\end{figure}
Fig.\ref{fig:2pcint} shows the hadron-hadron correlations in polar coordinates 
for trigger particle in $2<p_T^{trig}<4$ GeV/c and associated particles in $1<p_T^{asso}<2$ GeV/c in
Au+Au collisions at $\sqrt{s_{NN}}=200$ GeV, with the additional selection on 
the angle of the trigger particle with respect to the second order event plane.
Also in this case contributions from $v_2$ and $v_4(\Phi_2)$ are subtracted from correlations with ZYAM method.
In the left panel, the correlations are shown for the case of trigger particle in the out-of-plane region,
while the case of in-plane trigger particle is shown in the right panel.

The open symbols (in blue) show the correlations where the trigger particle is selected
in left side of event planes, while the closed symbols (in red) show the correlations
where the trigger particle is selected in right side of event plane.
The ellipse at the center (in green) represents the medium shape.
The red, blue, and black arrows indicate the direction of the trigger particle and of the event plane,
as well as orange arrows indicate the direction in which correlation yields are more pronounced. 

Focusing on the away side shape of left-triggered correlations from event plane,
the yield in longer path length tends to be larger than that in shorter path length side
in the case of trigger particle exiting out-of-plane, while, for the case of trigger particle
in the in-plane region, the yield in shorter path length side tends to be larger.
Out-of-plane triggered correlations are dominated by surface emission of particles,
however in-plane triggered correlations are overwhelmed by penetrating emissions.
These observations indicate a non-monotonic behavior of the away side yield as a function of
parton path length, which is inconsistent with the monotonic trends seen in high momenta pair correlations
shown in Fig.\ref{fig:2pchigh}.

This non-monotonic behavior in triggered correlations is preserved even including $v_3$ terms
to flow contribution subtraction. Since the correlation between second and third event planes
is weak and contributions from $v_3$ is almost independent of the angle between the trigger particle
and the second order event plane, and the same amplitude of contributions from $v_3$ is subtracted
in any correlations with different trigger particle angle relative to second order event planes.

\subsection{Two-particle correlations with $v_n$ contribution subtractions}
\begin{figure}[t]
\begin{center}
\includegraphics[scale=0.35]{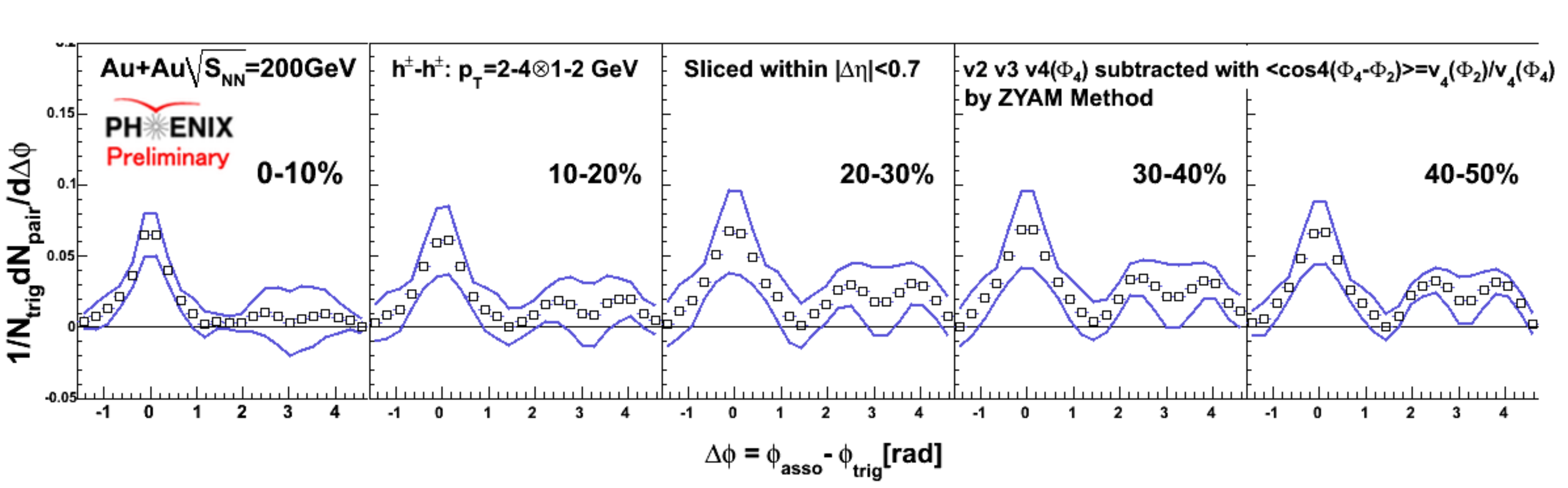}
\end{center}
\caption{
(Color online)
Azimuthal correlations of two charged hadrons in Au+Au collisions at $\sqrt{s_{NN}}=200$ GeV
in five centrality intervals in the range 0\%-50\%.
Transverse momentum ranges are 2-4 GeV/c for trigger particles and 1-2GeV/c for associated ones.
Contributions from $v_2$, $v_3$ and $v_4(\Phi_4)$ are subtracted with ZYAM method,
considering the correlations between experimentally observed second and fourth order event planes
defined as $\left<\cos{4(\Phi_2-\Phi_4)}\right>=v_4(\Phi_2)/v_4(\Phi_4)$.
}
\label{fig:vn2pc}
\end{figure}

Fig.\ref{fig:vn2pc} presents the hadron-hadron correlations for trigger particle in $2<p_T^{trig}<4$ GeV/c
and associated particles in $1<p_T^{asso}<2$ GeV/c without selecting on the angle between
the trigger particle and the event plane. The $v_2$, $v_3$ and $v_4$ are subtracted.
The subtraction of contributions from $v_3$ and $v_4$ reduce $\cos{3\Delta\phi}$ and $\cos{4\Delta\phi}$
components in correlation functions, resulting in the disappearance of the away side correlation yield
in most-central collisions, as seen in correlations with 
rapidity gap between pairs measured at the LHC\cite{Aamodt:2011prl107}\cite{Aad:2012prc86}.
However, away side yield and double humps remain even after the $v_n$ contribution subtraction in mid-central collisions.

The remaining amplitude of $v^{remain}_3$ which is estimated by the $\cos{3\Delta\phi}$ components
in flow subtracted correlations
is roughly equivalent to the amplitude of already subtracted $v^{subtracted}_3$. It is difficult to explain away side double hump
structure in mid-central collisions only with higher order flow harmonics $v_n$.

\section{Summary}
\label{Summary}
Two particle correlations with trigger particle selection relative to second order event plane have been
measured at high and intermediate transverse momenta pairs.
While away side yields in high momentum pair correlations
shows a monotonic suppression as a function of path length,
yields from two particle correlations at intermediate momentum show a non-monotonic behavior.
Two particle correlations without selection on the trigger particle angle relative to
the event plane and with subtraction of :the contribution from $v_n$ flow coefficients show the disappearance
 of double hump structure in most-central collisions.
However, the away side double hump structure is still present in mid-central collisions.
It is difficult to explain the remaining away side double humps in mid-central collisions in the context of higher order flow harmonics.





\bibliographystyle{elsarticle-num}
\bibliography{HP2012proceedings_todoroki}







\end{document}